\documentclass[showpacs,preprintnumbers,amsmath,amssymb,twocolumn]{revtex4}

\usepackage{graphicx}
\usepackage{dcolumn}

\usepackage{bm}

\begin{document}
%-----------------------------------------------
\title{General class of wormhole geometries in conformal Weyl gravity}
%-----------------------------------------------

\author{Francisco S. N. Lobo}
\email{francisco.lobo@port.ac.uk} \affiliation{Institute of
Gravitation \& Cosmology, University of Portsmouth, Portsmouth PO1
2EG, UK}
\email{flobo@cosmo.fis.fc.ul.pt} \affiliation{Centro de Astronomia
e Astrof\'{\i}sica da Universidade de Lisboa, Campo Grande, Ed. C8
1749-016 Lisboa, Portugal}

\date{\today}

%-----------------------------------------------
\begin{abstract}
%-----------------------------------------------

In this work, a general class of wormhole geometries in conformal
Weyl gravity is analyzed. A wide variety of exact solutions of
asymptotically flat spacetimes is found, in which the stress
energy tensor profile differs radically from its general
relativistic counterpart. In particular, a class of geometries is
constructed that satisfies the energy conditions in the throat
neighborhood, which is in clear contrast to the general
relativistic solutions.

%-----------------------------------------------
\end{abstract}
%-----------------------------------------------


%-----------------------------------------------
\maketitle
%-----------------------------------------------

%-----------------------------------------------
\section{Introduction}
%-----------------------------------------------

The Einstein field equation reflects the dynamics of general
relativity, and is formally obtained from the Einstein-Hilbert
action, $I_{\rm EH}=\int d^4x\,\sqrt{-g}\,R$, where $R$ is the
curvature scalar. However, the latter action can be generalized to
include other scalar invariants. An intriguing example is
conformal Weyl gravity, involving the following purely
gravitational sector of the action
\begin{eqnarray}\label{action1}
I_W&=&-\alpha \int d^4x
\sqrt{-g}\;C_{\mu\nu\alpha\beta}\,C^{\mu\nu\alpha\beta}
 \,,
\end{eqnarray}
where $C_{\mu\nu\alpha\beta}$ is the Weyl tensor, and $\alpha$ is
a dimensionless gravitational coupling constant. It was argued in
Ref. \cite{Mannheim:1988dj,Kazanas:1988qa} that in analogy to the
principle of local gauge invariance that severely restricts the
structure of possible Lorentz invariant actions in flat
spacetimes, then the principle of local conformal invariance is a
requisite invariance principle in curved spacetimes. The latter
principle requires that the gravitational action to remain
invariant under the conformal transformations
$g_{\mu\nu}(x)\rightarrow \Omega^2(x)g_{\mu\nu}(x)$. The conformal
Weyl tensor
\begin{equation}\label{Weyltensor}
C_{\mu\nu\alpha\beta}=R_{\mu\nu\alpha\beta}-g_{\mu[\alpha}R_{\beta]\nu}
+g_{\nu[\alpha}R_{\beta]\mu}
   +\frac{1}{3}Rg_{\mu[\alpha}g_{\beta]\nu}\,,
\end{equation}
also transforms as $C_{\mu\nu\alpha\beta}\rightarrow
\Omega^2(x)C_{\mu\nu\alpha\beta}$. The action $I_W$ is an
interesting theoretical construct, for instance, being a strictly
conformally invariant theory, particle masses may possibly arise
through the spontaneous symmetry breaking of the action
\cite{Mannheim:1988dj}.

Being a fourth order gravity theory, with respect to the
derivatives of the metric, finding exact solutions of the
gravitational field equations yields a formidable endeavor.
Nevertheless, the exact vacuum exterior solution for a static and
spherically symmetric spacetime in locally conformal invariant
Weyl gravity was found in Ref. \cite{Mannheim:1988dj}. The
solution contains the exterior Schwarzschild solution and provides
a potential explanation for observed galactic rotation curves
without the need for dark matter
\cite{Mannheim:1999bu,Edery:2001at}. The time-dependent
spherically symmetric solution was further explored in Ref.
\cite{Kazanas:1988qa}. The exact solutions to the
Reissner-Nordström problem associated with a static and
spherically symmetric point electric and/or magnetic charge
coupled to fourth-order conformal Weyl gravity were found
\cite{Mannheim:1990ya}. In addition to this, exact solutions
associated with the fourth-order Kerr and Kerr-Newman problems in
which a stationary and axially symmetric rotating system with or
without electric and/or magnetic charge is coupled to gravity,
were further explored \cite{Mannheim:1990ya}. The causal
structure, using Penrose diagrams, of the static spherically
symmetric vacuum solution to conformal Weyl gravity was also
investigated \cite{Edery:1998zi}.
New vacuum solutions were found using a covariant
$(2+2)$-decomposition of the field equation, which covers the
spherically and the plane symmetric space-times as special
subcases \cite{Dzhunushaliev:1999fy}. Exact topological black hole
solutions of conformal Weyl gravity, with negative, zero or
positive scalar curvature at infinity were also found
\cite{Klemm:1998kf}, the former generalizing the well-known
topological black holes in anti-de Sitter gravity.

The weak-field limit of conformal Weyl gravity for an arbitrary
spherically symmetric static distribution of matter in the
physical gauge with a constant scalar field was also analyzed
\cite{Barabash:2007ms}, and it was argued that the conformal
theory of gravity is inconsistent with the Solar System
observational data. In a cosmological context, exact analytical
solutions to conformal Weyl gravity for the matter and radiation
dominated eras, and the primordial nucleosynthesis process were
exhaustively analyzed. It was found that the cosmological models
are unlikely to reproduce the observational properties of our
Universe, as they fail to fulfill the observational constraints on
present cosmological parameters and on primordial light element
abundances~\cite{Elizondo:1994vh}. In Ref. \cite{Flanagan:2006ra}
it was also argued that in the limit of weak fields and
non-relativistic velocities the theory does not agree with the
predictions of general relativity, and is therefore ruled out by
Solar System observations. Nevertheless, in Ref.
\cite{Mannheim:2007ug}, it was counter-argued in the presence of
macroscopic long range scalar fields, the standard Schwarzschild
phenomenology is still recovered. To check the viability of Weyl
gravity, two additional classical tests of the theory, namely, the
deflection of light and time delay in the exterior of a static
spherically symmetric source were analyzed, and it was shown that
the parameters fit the experimental constraints
\cite{Edery:1997hu,Pireaux:2004xb}.

An interesting application of conformal Weyl gravity would be to
analyze traversable wormhole solutions in the theory. We emphasize
that an important and intriguing challenge in wormhole physics is
the quest to find a realistic matter source that will support
these exotic spacetimes. In classical general relativity,
wormholes are supported by exotic matter, which involves a stress
energy tensor that violates the null energy condition (NEC)
\cite{Morris:1988cz,Visser}. Note that the NEC is given by
$T_{\mu\nu}k^\mu k^\nu \geq 0$, where $k^\mu$ is {\it any} null
vector. Several candidates have been proposed in the literature,
amongst which we refer to solutions in higher dimensions, for
instance in Einstein-Gauss-Bonnet theory \cite{EGB1,EGB2},
wormholes on the brane \cite{braneWH1,braneWH2}; solutions in
Brans-Dicke theory \cite{Nandi:1997en}; wormhole solutions in
semi-classical gravity (see Ref. \cite{Garattini:2007ff} and
references therein); exact wormhole solutions using a more
systematic geometric approach were found \cite{Boehmer:2007rm};
and solutions supported by equations of state responsible for the
cosmic acceleration \cite{phantomWH}, etc (see Refs.
\cite{Lemos:2003jb,Lobo:2007zb} for more details and
\cite{Lobo:2007zb} for a recent review). In conformal Weyl
gravity, as the gravitational field equations differ radically
from the Einstein field equation, one would expect a wider class
of solutions. This is indeed the case, and the solutions found
contain interesting physical properties and characteristics,
amongst which we refer to a zero or positive radial pressure at
the throat, or more important the non-violation of the energy
conditions in the throat neighborhood, contrary to their general
relativistic counterparts.

This paper is outlined in the following manner: In Section
\ref{sec:II}, we outline the general formalism and the
gravitational field equations governing static and spherically
symmetric spacetimes in conformal Weyl gravity. In Section
\ref{sec:III}, we further explore specific wormhole solutions, and
finally, in Section \ref{sec:conclusion}, we conclude.

%-----------------------------------------------
\section{General formalism}\label{sec:II}
%-----------------------------------------------

\subsection{Gravitational field equations}

The metric used throughout this work, in curvature coordinates, is
given by
\begin{equation}\label{metric}
ds^2=-B(r)\,dt^2+A(r)\,dr^2+r^2\,\left(d\theta^2+\sin^2\theta
\,d\phi^2 \right)\,.
\end{equation}

The Weyl action, Eq. (\ref{action1}), may be simplified by noting
that the quantity
\begin{equation}\label{divergence}
\sqrt{-g}\left(R_{\mu\nu\alpha\beta}R^{\mu\nu\alpha\beta}
-4R_{\mu\nu}R^{\mu\nu}+R^2\right) \,,
\end{equation}
is a total divergence, and thus $I_W$ may be rewritten as
\begin{equation}\label{action}
I_W=-2\alpha \int
d^4x\sqrt{-g}\,\left(R_{\mu\nu}\,R^{\mu\nu}-\frac{1}{3}R^2\right)
     \,.
\end{equation}

Varying the action with respect to the metric $g_{\mu\nu}$
provides the following relationship
\begin{equation}\label{fieldeqs}
(-g)^{-1/2}\,g_{\mu\alpha}g_{\nu\beta}\frac{\delta I_W}{\delta
g_{\alpha\beta}}=-2\alpha\left[W_{\mu\nu}^{(2)}
-\frac{1}{3}W_{\mu\nu}^{(1)}\right]\,,
\end{equation}
with $W_{\mu\nu}^{(1)}$ and $W_{\mu\nu}^{(2)}$ given by
\begin{equation}\label{Weyl1}
W_{\mu\nu}^{(1)}=2g_{\mu\nu}R^{;\beta}{}_{;\beta}-2R_{;\mu\nu}
-2RR_{\mu\nu}+\frac{1}{2}g_{\mu\nu}R^2  \,,
\end{equation}
and
\begin{eqnarray}
W_{\mu\nu}^{(2)}&=&\frac{1}{2}g_{\mu\nu}R^{;\beta}{}_{;\beta}
+R_{\mu\nu}{}^{;\beta}{}_{;\beta}
-R_{\mu}{}^{\beta}{}_{;\nu\beta}-R_{\nu}{}^{\beta}{}_{;\mu\beta}
    \nonumber    \\
&&-2R_{\mu\beta}R_{\nu}{}^{\beta}+\frac{1}{2}g_{\mu\nu}R_{\alpha\beta}R^{\alpha\beta}\,,
    \label{Weyl2}
\end{eqnarray}
respectively.

The stress energy tensor is defined as
\begin{equation}\label{SET}
T_{\mu
\nu}=-\frac{2}{\sqrt{-g}}\frac{\delta(\sqrt{-g}\,L_m)}{\delta(g^{\mu\nu})}
\,,
\end{equation}
where $L_{m}$ is the Lagrangian density corresponding to matter.

The final gravitational field equation is given by
\begin{equation}\label{modfieldeq}
4\alpha\,W_{\mu\nu}=T_{\mu\nu}  \,,
\end{equation}
with $W_{\mu\nu}=W_{\mu\nu}^{(2)}-\frac{1}{3}W_{\mu\nu}^{(1)}$.
Both sides are symmetric, traceless and covariantly conserved.
Note that the intrinsic Newtonian constant that arises in the
Einstein-Hilbert action is absent.

Determining $W_{\mu\nu}$ from Eqs. (\ref{Weyl1}) and (\ref{Weyl2})
presents a formidable endeavor. However, one may use the fact that
for an arbitrary action $I=\int \sqrt{-g}\,d^4x\,L$, and using the
metric (\ref{metric}), the term $W^{rr}$ may be deduced from
\cite{Mannheim:1988dj}
\begin{eqnarray}
\sqrt{-g}\,W^{rr}=-\frac{1}{2\alpha}\frac{\delta I}{\delta
A}&=&\frac{\partial}{\partial
A}\left(\sqrt{-g}\,L\right)-\frac{\partial}{\partial
r}\left(\sqrt{-g}\,\frac{\partial L}{\partial A'}\right)
     \nonumber \\
&&+\frac{\partial^2}{\partial r^2}\left(\sqrt{-g}\,\frac{\partial
L}{\partial A''}\right) \,,
         \label{modLag}
\end{eqnarray}
where the prime denotes a derivative with respect to the radial
coordinate, $r$. Likewise for $W^{tt}$ from $\delta I/\delta B$,
etc. However, rather than use this method, which for calculational
purposes is rather intractable, the gravitational tensor
components $W^{tt}$ and $W^{\theta\theta}$ may be determined from
the Bianchi and trace identities, and given in terms of $W^{rr}$
\cite{Mannheim:1988dj}.

From the Bianchi identity,
\begin{equation}\label{BianchiIda}
W^{\mu\nu}{}_{;\mu}=(-g)^{-1/2}\left[(-g)^{1/2}\,W^{\mu\nu}\right]_{,\mu}
+\Gamma^\nu_{\mu\lambda}\,W^{\mu\lambda}=0 \,,
\end{equation}
one obtains the following relationship
\begin{equation}\label{BianchiId}
{\cal
D}\,W^{rr}+\frac{B'}{2A}\,W^{tt}-\frac{2r}{A}W^{\theta\theta}=0
\,,
\end{equation}
where we have defined
\begin{equation}\label{defD}
{\cal D}=\frac{\partial}{\partial
r}+\frac{2}{r}+\frac{A'}{A}+\frac{B'}{2B} \,,
\end{equation}
for notational simplicity.

From the trace identity, $W^{\mu}{}_{\mu}=0$, one obtains the
following relationship
\begin{equation}\label{traceId}
-B\,W^{tt}+A\,W^{rr}+2r^2\,W^{\theta\theta}=0 \,.
\end{equation}

Finally, using Eqs. (\ref{BianchiId}) and (\ref{traceId}), the
gravitational tensor components $W^{tt}$ and $W^{\theta\theta}$
are related to $W^{rr}$ through the following expressions
\begin{eqnarray}
W^{tt}&=&\frac{A}{B-B'r/2}\,\left(1+r\,{\cal D}\right)\,W^{rr}\,,
    \label{defWtt}
     \\
W^{\theta\theta}&=&\frac{A}{4r\left(B-B'r/2\right)}\,\left(B'+2B\,{\cal
D}\right)\,W^{rr}\,, \label{defWthth}
\end{eqnarray}
so that all the information is contained in the $W^{rr}$ term.

The stress-energy tensor components, through the gravitational
field equation, are given by
\begin{eqnarray}
\rho=-4\alpha\,W^{t}{}_{t}\,,\quad
p_r=4\alpha\,W^{r}{}_{r}\,,\quad
p_t=4\alpha\,W^{\theta}{}_{\theta}\,,
         \label{stress}
\end{eqnarray}
in which $\rho(r)$ is the energy density, $p_r(r)$ is the radial
pressure, and $p_t(r)$ is the lateral pressure measured in the
orthogonal direction to the radial direction. Note that in
conformal Weyl gravity, the stress energy tensor components are
constrained through the trace identity, i.e., $-\rho+p_r+2p_t=0$.

Although extremely lengthy, we present the relevant gravitational
terms, namely, $W^{r}{}_{r}$ and $W^{t}{}_{t}$, which will be used
extensively throughout this work:
\begin{widetext}
\begin{eqnarray}\label{expWrr}
W^{r}{}_{r}&=&\Big\{\big[4A^2B^2(2B'B'''-B''^2)
-4ABB''\left(3AB'^2+2A'BB'\right)+7A^2B'^4+6AA'BB'^3+
B^2B'^2\left(7A'^2-4AA''\right)\big]\,r^4
     \nonumber \\
&&+\big[-16A^2B^3B'''+16AB^2B''(3AB'+A'B)-20A^2BB'^3
-16AA'B^2B'^2+4B^3B'(4AA''-7A'^2)\big]r^3
     \nonumber    \\
&&+\big[-4A^2B^2(8BB''+B'^2)+8AA'B^3B' +4B^4(7A'^2-4AA'')\big]r^2
    \nonumber    \\
&&     +32A^2B^3B'r
    +16A^2B^4(A^2-1) \Big\}/(48A^4\,B^4r^4) \,,
\end{eqnarray}
%\end{widetext}
and
%\begin{widetext}
\begin{eqnarray}
W^{t}{}_{t}&=&\Big\{\Big[16A^2B^3B''''-48AB^2(AB'+A'B)B'''-36A^2B^2(B'')^2
+4ABB'(29AB'+27A'B)B''
     \nonumber   \\
   && -4B^3B''(8AA''-19A'^2)-49A^2B'^4-58AA'BB'^3+(24AA''-57A'^2)B^2B'^2
   \nonumber  \\
   &&+4\left(-2AA'''+13A'A''-14\frac{A'^2}{A}\right)B^3B'\Big]r^4+
   \Big[64A^2B^3B'''-104AB^2(AB'+A'B)B''
   \nonumber     \\
   &&4ABB'^2(11AB'+12A'B)-4(6AA''-13A'^2)B^3B'+16AA'''B^4-104AA''B^4
   +112\frac{A'^3}{A}B^4\Big]r^3
   \nonumber    \\
   &&\hspace{-0.75cm}\Big[20AB^2B'(AB'+2A'B)+16AA''B^4-28A'^2A''B^4\Big]r^2
   %  \nonumber    \\
     +32A^2B^3B'r
    +16A^2B^4(A^2-1) \Big\}/(48A^4\,B^4r^4) \,,
    \label{expWtt}
\end{eqnarray}

\end{widetext}
respectively. The term $W^{\theta}{}_{\theta}$ may be given by Eq.
(\ref{defWthth}), or simply using the trace identity, i.e.,
$W^{\theta}{}_{\theta}=-(W^{t}{}_{t}+W^{r}{}_{r})/2$, through Eqs.
(\ref{expWrr}) and (\ref{expWtt}).

\subsection{Energy conditions}

In this work we are interested in deducing exact solutions of
traversable wormholes in conformal Weyl gravity and, therefore, a
fundamental point is the energy condition violations. However, a
subtle issue needs to be pointed out in this respect. Note that
the energy conditions arise when one refers back to the
Raychaudhuri equation for the expansion where a term
$R_{\mu\nu}k^\mu k^\nu$ appears, and $k^\mu$ is a null vector. The
positivity of this quantity ensures that geodesic congruences
focus within a finite value of the parameter labelling points on
the geodesics. However, in general relativity, through the
Einstein field equation one can write the above condition in terms
of the stress energy tensor $T_{\mu\nu}$, and consequently one
ends up with the null energy condition given by $T_{\mu\nu}k^\mu
k^\nu \ge 0$. In any other theory of gravity, one would require to
know how one can replace $R_{\mu\nu}$ using the corresponding
field equations and hence using matter stresses. In particular, in
a theory where we still have an Einstein-Hilbert term, the task of
evaluating $R_{\mu\nu}k^\mu k^\nu$ is trivial. However, in the
conformal Weyl gravity under consideration, things are not so
straightforward.

To this effect, one may rewrite the gravitational field equation
(\ref{modfieldeq}) in terms of the Einstein tensor, in an
analogous form to the Einstein field equation, given by
\begin{equation}\label{modfieldeq2}
G_{\mu\nu}\equiv
R_{\mu\nu}-\frac{1}{2}g_{\mu\nu}R=\frac{1}{4\alpha}T^{{\rm
eff}}_{\mu\nu}\,,
\end{equation}
where the effective stress energy tensor is given by $T^{{\rm
eff}}_{\mu\nu}=T^{(m)}_{\mu\nu}+T^{(W)}_{\mu\nu}$. Note that this
relationship differs fundamentally from the Einstein field
equation, as one is considering a dimensionless gravitational
coupling constant $\alpha$, contrary to the Newtonian
gravitational constant $G$. Nevertheless, the gravitational field
equation written in this form proves extremely useful in deducing
a definition of the null energy condition, in terms of the
effective stress energy tensor, from the Raychaudhuri expansion
term $R_{\mu\nu}k^\mu k^\nu$.

The first term, i.e., $T^{(m)}_{\mu\nu}$, in the effective stress
energy tensor, is defined in terms of the matter stress energy
tensor, Eq. (\ref{SET}), and is given by
\begin{equation}
T^{(m)}_{\mu\nu}\equiv \frac{3}{2R}\,T_{\mu\nu}\,,
\end{equation}
where $R$ is the curvature scalar.

The second term $T^{(W)}_{\mu\nu}$ may be denoted as the curvature
Weyl stress energy tensor, and is provided by
\begin{equation}
T^{(W)}_{\mu\nu}\equiv
-\frac{6\alpha}{R}\,\overline{W}_{\mu\nu}\,,
\end{equation}
with the tensor $\overline{W}_{\mu\nu}$ defined as
\begin{eqnarray}
\overline{W}_{\mu\nu}&=&-\frac{1}{6}g_{\mu\nu}R^{;\beta}{}_{;\beta}
+R_{\mu\nu}{}^{;\beta}{}_{;\beta}
-R_{\mu}{}^{\beta}{}_{;\nu\beta}-R_{\nu}{}^{\beta}{}_{;\mu\beta}
    \nonumber    \\
&&\hspace{-1.2cm}-2R_{\mu\beta}R_{\nu}{}^{\beta}+\frac{1}{2}g_{\mu\nu}
R_{\alpha\beta}R^{\alpha\beta}+\frac{2}{3}R_{;\mu\nu}
+\frac{1}{6}g_{\mu\nu} R^2\,.
    \label{Weylmod}
\end{eqnarray}

Note that the gravitational field equation (\ref{modfieldeq2})
imposes interesting conservation equations. Through the the
Bianchi identities, $G^{\mu\nu}{}_{;\nu}=0$ and the conservation
of the stress energy tensor $T^{\mu\nu}{}_{;\nu}=0$, which can
also be verified from the diffeomorphism invariance of the matter
part of the action, one verifies the following conservation law
\begin{equation}
T^{(W)\mu\nu}{}_{;\nu}=\frac{3}{2R^2}\, T^{\mu\nu}R_{,\nu}\,.
\end{equation}

Now the positivity condition, $R_{\mu\nu}k^\mu k^\nu \geq 0$, in
the Raychaudhuri equation provides the following form for the null
energy condition $T^{{\rm eff}}_{\mu\nu} k^\mu k^\nu\geq 0$,
through the modified gravitational field equation
(\ref{modfieldeq2}). For this case, in principle, one may impose
that the matter stress energy tensor satisfies the energy
conditions and the respective violations arise from the Weyl
curvature term $T^{(W)}_{\mu\nu}$, in analogy to the case carried
out in Ref. \cite{braneWH2}. Although this analysis is an
interesting avenue to study, we consider an alternative approach
which is described below.

Another approach to the energy conditions considers in taking the
condition $T_{\mu\nu} k^\mu k^\nu\ge 0$ at face value. Note that
this is useful as using local Lorentz transformations it is
possible to show that the above condition implies that the energy
density is positive in all local frames of reference. However, if
the theory of gravity is chosen to be non-Einsteinian, then the
assumption of the above condition does not necessarily imply
focusing of geodesics. The focusing criterion is different and
will follow from the nature of $R_{\mu\nu} k^\mu k^\nu$. In the
next section, we consider this latter approach to the energy
conditions, which provides interesting results.

%-----------------------------------------------------------------
\section{Traversable wormholes in conformal Weyl gravity}
\label{sec:III}
%-----------------------------------------------------------------

In this section, we consider the equations of structure for
traversable wormholes in conformal Weyl gravity. For this, it is
convenient to express the metric in a more familiar form
\cite{Morris:1988cz,Visser}, given by
\begin{equation}
ds^2=-e ^{2\Phi(r)}\,dt^2+\frac{dr^2}{1- b(r)/r}+r^2 \,(d\theta
^2+\sin ^2{\theta} \, d\phi ^2) \label{metricwormhole}\,,
\end{equation}
where $\Phi(r)$ and $b(r)$ are arbitrary functions of the radial
coordinate, $r$, denoted as the redshift function and the form
function, respectively \cite{Morris:1988cz}. The radial coordinate
has a range that increases from a minimum value at $r_0$,
corresponding to the wormhole throat, to $\infty$.

To avoid the presence of event horizons, $\Phi(r)$ is imposed to
be finite throughout the coordinate range. At the throat $r_0$,
one has $b(r_0)=r_0$, which implies that $A(r_0)\rightarrow
\infty$. A fundamental condition is the flaring-out condition
given by $(b'r-b)/b^2<0$, which is provided by the mathematics of
embedding~\cite{Morris:1988cz,Visser}.

In analogy to their general relativistic counterparts, one may
consider asymptotically flat spacetimes. However, it is also
possible to match the interior wormhole solution to the unique
vacuum solution given by
\begin{equation}\label{vacuum}
B(r)=A^{-1}(r)=1-\frac{\beta(2-3\beta\gamma)}{r}-3\beta\gamma
+\gamma r-kr^2\,,
\end{equation}
where $\beta$, $\gamma$ and $k$ are constants of integration
\cite{Mannheim:1988dj,Kazanas:1988qa}. Note that the general
relativistic Schwarzschild solution is parameterized by $\beta$.
The constant $k$ characterizes a background de Sitter spacetime,
although the metric fields (\ref{vacuum}) in Weyl gravity
correspond to a vacuum solution. The integration constant $\gamma$
measures departures from the respective solution in classical
general relativity. Therefore, it is possible to have a cosmology
that admits a de Sitter solution without a cosmological constant
\cite{Mannheim:1989jh}. This latter term vanishes identically due
to the conformal invariance of the theory. Thus, conformal Weyl
gravity naturally avoids the theoretical--observational value
discrepancy of the cosmological constant.

In the analysis that follows, we consider that the factor that
appears in the gravitational field equation be equal to unity,
i.e., $4\alpha=1$, for notational and computational simplicity.

%-----------------------------------------------------
\subsection{Specific case: constant redshift function}
%-----------------------------------------------------

A particularly interesting case are the solutions with a constant
redshift function, $\Phi'=0$. Without a loss of generality one may
impose $\Phi=0$, which is equivalent to considering $B=1$. This
specific case simplifies the field equations significantly, and
provide particularly intriguing solutions, which differ from their
general relativistic counterparts. This is due to the fact that
the fourth order gravitational field equation in conformal Weyl
gravity differs from the general relativistic Einstein field
equation.

The energy density and radial pressure, taking into account Eqs.
(\ref{expWrr}) and (\ref{expWtt}), reduce to
\begin{eqnarray}
\rho&=&-\frac{1}{6r^6}\Bigg\{2r^2\left(b'''r^2-2b''r
+2b'\right)\left(1-\frac{b}{r}\right)
     \nonumber   \\
&&+\left[b''r^2+\frac{5}{2}\left(b-b'r\right)\right]
\left(b-b'r\right)+2b^2\Bigg\}\,,
   \label{nullPhirho}
\\
%\end{eqnarray}
%
%\begin{eqnarray}
p_r&=&\frac{1}{6r^6}\Bigg\{2r^2\left(-b''r+2b'\right)\left(1-\frac{b}{r}\right)
     \nonumber   \\
&&+\frac{b'r}{2}\left(b-b'r\right)+\frac{b}{2}(3b+b'r)\Bigg\}\,,
          \label{nullPhipr}
\end{eqnarray}
respectively. The NEC is given by $T_{\mu\nu}k^\mu k^\nu \geq 0$,
as mentioned in the Introduction, and for a diagonal stress energy
tensor takes the form $\rho+p_r\geq 0$. For the present case, the
NEC is given by
\begin{eqnarray}
\rho+p_r&=&-\frac{1}{6r^6}\Bigg\{2r^3\left(b'''r-b''\right)\left(1-\frac{b}{r}\right)
     \nonumber   \\
&&+\left[b''r^2+3(b-b'r)\right]\left(b-b'r\right)\Bigg\}\,.
    \label{nullPhiNEC}
\end{eqnarray}

To verify the non-violation of the NEC at the throat, Eq.
(\ref{nullPhiNEC}) imposes the following inequality
\begin{equation}\label{nullPhiNEC2}
b''r_0\leq 3(b'-1) \,,
\end{equation}
where the flaring-out condition evaluated at the throat has been
taken into account, i.e., $b'(r_0)<1$. We consider next specific
choices for the form function.

%-----------------------------------------------
\subsubsection{Form function: $b(r)=r_0$}
%-----------------------------------------------

For this case, the stress energy tensor components are given by
\begin{equation}
\rho=-\frac{3r_0^2}{4r^6}\,,\qquad  p_r=\frac{r_0^2}{4r^6}\,.
\end{equation}
Note that in this simple case, one already obtains a solution that
deviates from the general relativistic counterpart, in that the
radial pressure is positive at the throat. Recall that in general
relativity the radial pressure is always negative at the throat,
implying the necessity of a radial tension to maintain the throat
open. In addition to this, we recall that for the specific case of
$b(r)=r_0$, the energy density in general relativity is zero,
whilst in conformal Weyl gravity it is negative.

The NEC is provided by
\begin{equation}
\rho+p_r=-\frac{r_0^2}{2r^6}\,,
\end{equation}
which shows that the NEC is violated throughout the spacetime.

%-----------------------------------------------
\subsubsection{Form function: $b(r)=r_0^2/r$}
%-----------------------------------------------

The specific case of $b(r)=r_0^2/r$ corresponds to a negative
energy density in general relativity. Equations
(\ref{nullPhirho})-(\ref{nullPhipr}) provide the following stress
energy tensor scenario
\begin{equation}
\rho=\frac{4(3r^2-5r_0^2)r_0^2}{3r^8}\,,\qquad
p_r=-\frac{4(r^2-r_0^2)r_0^2}{3r^8}\,.
\end{equation}
The energy density negative in the range $r_0\leq r
<\sqrt{5/3}r_0$. This example also differs from its general
relativistic counterpart in that the radial pressure is zero at
the throat.

The NEC is provided by
\begin{equation}
\rho+p_r=-\frac{8(r^2-2r_0^2)r_0^2}{3r^8}\,.
\end{equation}
which shows that the NEC is violated for $r_0\leq r <
\sqrt{2}r_0$.

%------------------------------------------------------------
\subsubsection{Form function: $b(r)=r_0+\gamma r_0\left(1-r_0/r\right)$}
%------------------------------------------------------------

\begin{figure}[h]
  \centering
  \includegraphics[width=2.6in]{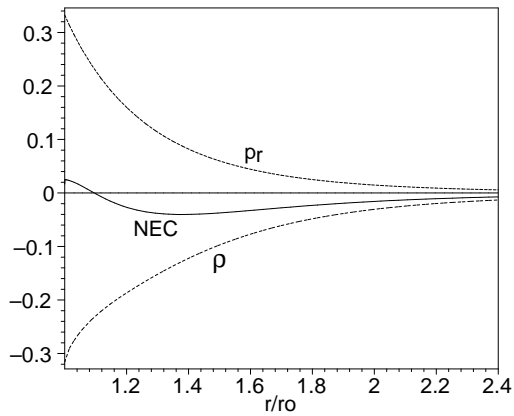}
  \caption{The energy density, radial pressure and NEC profile for
  the specific case of $\Phi'(r)=0$ and $b(r)=r_0+\gamma
  r_0(1-r_0/r)$ for $\gamma=0.9$. The energy density is negative,
  the radial pressure positive; and the NEC is satisfied at the
  throat neighborhood. In particular, at the throat the NEC is
  satisfied in the range of $0.6\leq \gamma < 1$.
  See the text for details.}
  \label{plot:gamma1}
\end{figure}

The specific choice of
\begin{equation}\label{form}
b(r)=r_0+\gamma r_0\left(1-\frac{r_0}{r}\right)\,,
\end{equation}
where $0<\gamma<1$, is particularly interesting. The stress energy
tensor components are somewhat lengthy, so that the respective
profile of the energy density, radial pressure and the NEC are
depicted in Fig. \ref{plot:gamma1}.

It is interesting to note that for this specific case the NEC
evaluated at the throat is given by
\begin{equation}
\left(\rho+p_r\right)\big|_{r_0}=-\frac{5\gamma^2-8\gamma+3}{6r_0^4}\,.
\end{equation}
This choice does indeed eliminate the need for the violation of
the NEC, in the interval $0.6\leq\gamma<1$. Note that this is
consistent with the general condition given by inequality
(\ref{nullPhiNEC2}). Nevertheless, the energy density is negative
throughout the spacetime, which violates the weak energy condition
(WEC). The WEC, $T_{\mu\nu}U^\mu U^\nu \geq 0$, where $U^\mu$ is a
timelike vector, implies $\rho\geq 0$ and $\rho+p_r\geq 0$. Note
that the radial pressure is positive as depicted in Fig.
\ref{plot:gamma1}. The specific case of $\gamma=0.9$ has been used
in the figure, which may be considered as a representative for
this specific case.

\begin{figure}[h]
  \centering
  \includegraphics[width=2.6in]{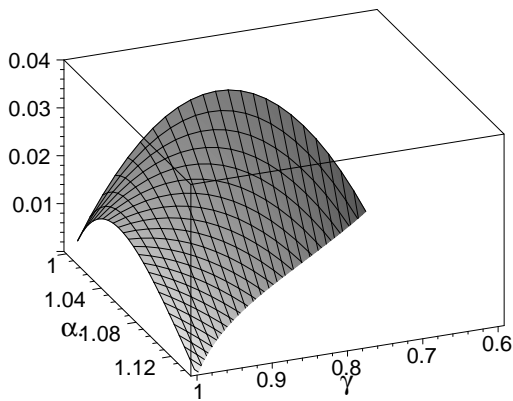}
  \caption{The NEC profile, with $\rho+p_r\geq 0$, for
  the specific case of $\Phi'(r)=0$ and $b(r)=r_0+\gamma
  r_0(1-r_0/r)$. We have defined $\alpha=r/r_0$.
  The NEC is satisfied at the throat in the range
  of $0.6\leq \gamma < 1$. One verifies, qualitatively,
  that the NEC is satisfied for high values of $\gamma$ and
  low values of $r$, i.e., as $r$ increases, then $\gamma$
  tends to its limiting value of $1$.}
  \label{plot:gamma2}
\end{figure}

The qualitative behavior of the NEC is depicted in Fig.
\ref{plot:gamma2}. Note that the NEC is satisfied for high values
of $\gamma$ and low values of $r$. In particular, the NEC is
satisfied for increasing values of $r$, as $\gamma$ tends to its
limiting value of $1$.

%-----------------------------------------------
\subsection{Specific case: $\Phi(r)=r_0/r$}
%-----------------------------------------------

For this case the stress energy tensor components are extremely
lengthy, so that they are also depicted in the respective plots
for the specific choices of the form function, considered below.

\subsubsection{Form function: $b(r)=r_0$}

\begin{figure}[h]
  \centering
  \includegraphics[width=2.6in]{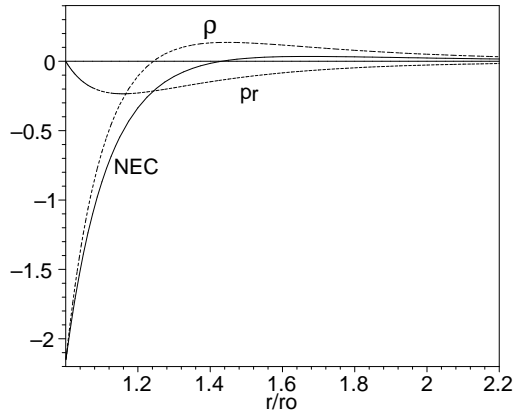}
  \caption{The energy density, radial pressure and NEC profile for
  the specific case of $\Phi(r)=r_0/r$ and $b(r)=r_0$.
  The radial pressure is zero at the throat; the energy
  density is negative and the NEC is violated in the throat's
  neighborhood.} \label{plot:Phi1}
\end{figure}

The energy density, radial pressure and NEC are depicted in Fig.
\ref{plot:Phi1}. Note that the radial pressure is zero at the
throat, and then remains negative throughout the coordinate range.
The energy density and the NEC are negative in the throat
neighborhood.

\subsubsection{Form function: $b(r)=r_0^2/r$}

\begin{figure}[h]
  \centering
  \includegraphics[width=2.6in]{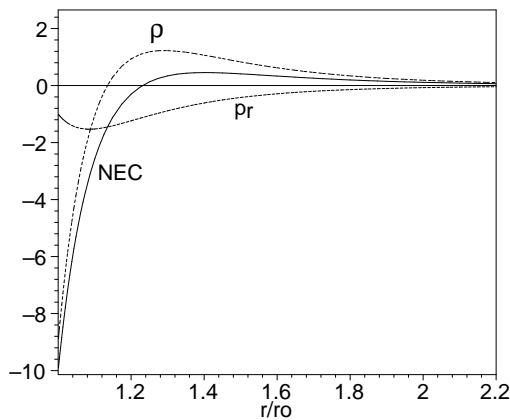}
  \caption{The energy density, radial pressure and NEC profile for
  the specific case of $\Phi(r)=r_0/r$ and $b(r)=r_0^2/r$.
  The radial pressure is negative throughout the spacetime; the energy
  density is negative and the NEC is violated in the neighborhood
  of the throat.} \label{plot:Phi1b}
\end{figure}

The energy density, radial pressure and NEC are depicted in Fig.
\ref{plot:Phi1b}. This choice is qualitatively analogous to the
previous case, except that the radial pressure is negative at the
throat.

\subsection{Specific case: $\Phi(r)=-r_0/r$ and $b(r)=r_0$}

\begin{figure}[h]
  \centering
  \includegraphics[width=2.6in]{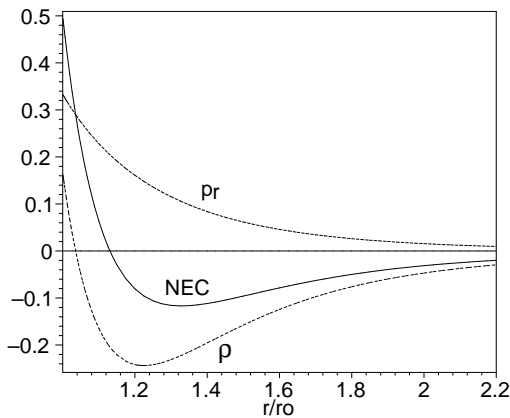}
  \caption{The energy density, radial pressure and NEC profile for
  the specific case of $\Phi(r)=-r_0/r$ and $b(r)=r_0$. The radial
  pressure is positive throughout.
  The energy density and the NEC are positive in
  the throat's neighborhood. Consequently, this example shows that one
may, in principle, construct a class of traversable wormholes,
within the context of conformal Weyl gravity, that satisfies all
of the energy conditions, in the vicinity of the throat.}
\label{plot:Phi2}
\end{figure}

This specific example is a considerable improvement to the
solutions considered above. The energy density, radial pressure
and NEC profile are depicted in Fig. \ref{plot:Phi2}. Note that
the pressure is always positive, and the energy density and NEC
are also positive in the neighborhood of the throat, thus
satisfying all of the energy conditions.

The profile for the specific case of $b(r)=r_0^2/r$ is
qualitatively analogous to this case. One may then match these
solutions to the exterior vacuum given by Eq. (\ref{vacuum}), at a
junction surface $a_0$, in which the energy conditions are
satisfied in the interval $r_0\leq r\leq a_0$. This shows that one
may, in principle, construct a class of traversable wormholes,
within the context of conformal Weyl gravity, that satisfies all
of the energy conditions, contrary to their general relativistic
counterparts.

%\bigskip

%-----------------------------------------------
\section{Conclusion}\label{sec:conclusion}
%-----------------------------------------------

In general relativity, the null energy condition violation is a
fundamental ingredient of static traversable wormholes. Despite
this fact, it was shown that for time-dependent wormhole solutions
the null energy condition and the weak energy condition can be
avoided in certain regions and for specific intervals of time at
the throat \cite{dynamicWH}. Nevertheless, in certain alternative
theories to general relativity, taking into account the modified
Einstein field equation, one may impose in principle that the
stress energy tensor threading the wormhole satisfies the NEC.
However, we emphasize that the latter is necessarily violated by
an effective total stress energy tensor. This is the case, for
instance, in braneworld wormhole solutions, where the matter
confined on the brane satisfies the energy conditions, and it is
the local high-energy bulk effects and nonlocal corrections from
the Weyl curvature in the bulk that induce a NEC violating
signature on the brane \cite{braneWH2}. Another particularly
interesting example is in the context of the $D$-dimensional
Einstein-Gauss-Bonnet theory of gravitation \cite{EGB1}, where it
was shown that the weak energy condition can be satisfied
depending on the parameters of the theory.

In this work, a general class of wormhole geometries in conformal
Weyl gravity was analyzed. In conformal Weyl gravity, as the
fourth order gravitational field equations differ radically from
the Einstein field equation, one would expect a wider class of
solutions. This is indeed the case, in which the stress energy
tensor profile differs radically from its general relativistic
counterpart, amongst which we may refer to a zero or positive
radial pressure at the throat, or at a more fundamental level, the
non-violation of the energy conditions in the throat neighborhood,
which is in clear contrast to the classical general relativistic
static wormhole solutions. Note that as for their general
relativistic counterparts, these Weyl variations have far-reaching
physical implications, namely apart from being used for
interstellar shortcuts, and being multiply-connected spacetimes an
absurdly advanced civilization may convert them into
time-machines~\cite{mty,Visser,Kluwer}, probably implying the
violation of causality.

%-----------------------------------
\section*{Acknowledgements}
%-----------------------------------
I thank Demos Kazanas for extremely stimulating discussions. This
work was funded by Funda\c{c}\~{a}o para a Ci\^{e}ncia e a
Tecnologia (FCT)--Portugal through the grant SFRH/BPD/26269/2006.

%-----------------------------------------------

%-----------------------------------------------

%-----------------------------------------------
\end{document}